\documentclass[pdftex,twocolumn,epjc3]{svjour3}          

\RequirePackage[T1]{fontenc}

\RequirePackage{graphicx}
\RequirePackage{mathptmx}      
\RequirePackage{flushend}
\usepackage{wasysym}
\usepackage{breqn}
\RequirePackage[numbers,sort&compress]{natbib}
\RequirePackage[colorlinks,citecolor=blue,urlcolor=blue,linkcolor=blue]{hyperref}

\begin{document}

\title{Wormholes in viable $f(R)$ modified theories of gravity and Weak Energy Condition}

\author{Petar Pavlovic\thanksref{e1,addr1}
        \and
        Marko Sossich\thanksref{e2,addr2} 
}

\thankstext{e1}{e-mail: petar.pavlovic@desy.de}
\thankstext{e2}{e-mail: marko.sossich@net.hr}

\institute{II. Institut f\"{u}r Theoretische Physik,
Universit\"{a}t Hamburg, Luruper Chaussee 149, 22761 Hamburg, Germany\label{addr1}
          \and
          Department of Physics, Faculty of Electrical Engineering and Computing, University of Zagreb,
          Unska 3, HR-10 000 Zagreb, Croatia\label{addr2}
}

\maketitle

\begin{abstract}
In this work wormholes in viable $f(R)$ gravity models are analysed. We are 
 interested in exact solutions for stress-energy 
 tensor components depending on different shape and redshift functions.
 Several solutions of gravitational equations for different $f(R)$ models are examined. 
 Found solutions imply no need for exotic material, while this need is implied
  in the standard general theory of relativity. A simple expression for WEC violation near the throat is derived and analysed. High curvature regime is also
  discussed, as well as the question of the highest possible values of the Ricci scalar for which the WEC is
  not violated near the throat, and corresponding functions are calculated for several models.  
  The approach  here  differs from the one that has been common since no 
 additional assumptions to simplify the equations have been made, and functions in
  $f(R)$ models are not
 considered to be arbitrary functions, but rather a feature of the theory that 
 has to be evaluated on the basis of consistency with observations for the Solar System and cosmological evolution. 
 Therefore in this work we show that existence of wormholes without exotic matter is not only possible in simple 
 arbitrary $f(R)$ models, but also in models that are in accordance with empirical data.
\end{abstract}

\section{Introduction}
Modified $f(R)$ gravity represents a possible alternative to Einstein's theory of general relativity 
which has received increased attention in the last decade. It is based on a generalization of the Einstein field equations 
that comes as a result of replacing the Ricci scalar curvature, $R$, with an arbitrary function of the scalar curvature, $f(R)$, 
in the gravitational Lagrangian density. One of the main reasons for increased interest in modified gravity theories comes 
from the possibility of explaining accelerating expansion of the universe, that has basically been confirmed by observations 
from type Ia supernovae \cite{perlmutter, grant, riess}, but also from other cosmological observations such as those from
large scale structure \cite{eisenstein}, and cosmic microwave background radiation \cite{spergel}. An important feature of $f(R)$ 
gravity is that in its framework, unlike in $\Lambda$ CDM cosmology based on the standard general relativity, there is no need for postulating dark energy or introducing any 
kind of new scalar or spinor field to explain the accelerated expansion \cite{faraoni, nojiri}. 
The action for $f(R)$ theories is given by
\begin{equation}
 S=\frac{1}{2k} \int \sqrt{-g}f(R) d^{4}x + S_{MAT},
 \label{akcija}
 \end{equation}  
  where $k=8 \pi G$, $g$ is a determinant of the metric, and $S_{MAT}$ is the matter action. 
Depending on the assumptions taken in the variational procedure starting from 
(\ref{akcija}) we can make a distinction between metric, Palatini and metric-affine 
formalism \cite{buchadahl, liberati}. In metric formalism we proceed from the 
assumption that the connection is dependent on the metric, namely that it is given by the 
Christoffel connection. In Palatini formalism the connection is treated  
independent of the metric and it is also assumed that the matter part of the action is 
not  
dependent on the connection. Finally, in metric-affine formalism the matter part of the action now depends on the 
connection which is metric independent. In this work we will use the metric approach which is the simplest of the above 
mentioned and also usually used in literature. Using this approach and varying the action (\ref{akcija}) with respect to the metric we obtain the 
following modified field equations
\begin{equation}
R_{\mu \nu}f_{R}(R) - \frac{1}{2}
g_{\mu \nu}f(R)-(\nabla_{\mu}\nabla_{\nu}-
g_{\mu \nu}\Square)f_{R}(R) = k T_{\mu \nu} ,
\label{modificirana einsteinova}
\end{equation}
where
\begin{equation}
T_{\mu \nu}=\frac{-2}{\sqrt{-g}} \frac{\delta S_{MAT}}
{\delta g^{\mu \nu}} ,
\end{equation}
 $f_{R}=df(R)/dR$, and we will use an analogous notation for higher derivatives of $f(R)$. Adopting the standard definition: $G_{\mu \nu}=R_{\mu \nu} - 1/2 Rg_{\mu \nu}$, and after 
 some mathematical manipulations, we can obtain a following equation for the Einstein tensor from (\ref{modificirana einsteinova}) 
\begin{dmath}
G_{\mu \nu}= \frac{1}{f_{R}}\lbrace f_{RR}\nabla_{\mu}\nabla_{\nu}R + f_{RRR}(\nabla_{\mu}R)(\nabla_{\nu}R)  \\ - \frac{g_{\mu \nu}}{6}(R f_{R}+ f + 16 \pi GT)+ 8 \pi G T_{\mu \nu}\rbrace, 
\label{einstein tensor}
\end{dmath}
where $T$ is the trace of the stress-energy tensor. In this work we have analysed   wormhole solutions in the framework of viable metric $f(R)$ gravity models which do not violate standard energy conditions.
Wormholes are hypothetical tunnels with a throat that connects two asymptotically flat regions of spacetime. In Einstein's general relativity, a construction
of a wormhole is possible only by the use of exotic matter i.e. matter that violates usual energy conditions \cite{Thorne, hoc, Visser}. The matter threading the wormhole is 
usually described by the perfect anisotropic
fluid $T_{\mu\nu}=diag(\rho,p_{r},p_{t},p_{t})$. It can be shown that existence of a wormhole in General relativity implies the condition $\rho + p_{r}<0$ and 
according to \cite{Thorne} we 
shall call the material with this property exotic. This violates the Weak Energy Condition (WEC), which is given by $T_{\mu\nu}k^{\mu}k^{\nu}\geq0$ for any timelike 
vector $k^{\mu}$ \cite{Hawking}.
The Weak Energy Condition expresses constraints on a possible matter behaviour in order to guarantee some usual properties, such as positive energy density. On the other hand, 
it was reported that static spherically symmetric wormhole can be supported by a phantom energy \cite{sushov}. The general question of the WEC violation in modefied gravity 
still remains open \cite{parikh}. It was shown in \cite{harkoilobo} that in modified gravity with the field equations of the form
\begin{equation}
g_{1}(\psi^{i})(G_{\mu \nu} + H_{\mu \nu}) - g_{2}(\psi^{j})T_{\mu \nu}=k T_{\mu \nu},
\end{equation}
where $H_{\mu \nu}$ is an additional geometrical structure, $g_{i}(\psi^{j})$ multiplicative factors, and $\psi^{j}$ curvature invariants of gravitational field, normal matter threading the wormhole can satisfy the WEC if it fulfils following relationship
\begin{equation}
\frac {g_{1}(\psi^{i})}{k+g_{2}(\psi^{j})}(G_{\mu \nu} + H_{\mu \nu})k^{\mu} k^{\nu} \geq 0 .
\end{equation} 
In modified $f(R)$ theories of gravity, wormholes can be supported by ordinary matter \cite{lobo, Oliveira, nasa}. Therefore, while interested in WEC non-violation we 
are exploring solutions that satisfy $\rho\geq0$ and $\rho + p_{r}\geq0$ \cite{lobo}.
Our aim is to analyse, without any additional assumptions, possible wormhole solutions in different viable recently proposed $f(R)$ models that do not imply
the existence of exotic material. In Section 2. we present wormhole geometry, effective field equations and derive suitable expressions for the WEC non violation near the throat.
In Section 3. we present
and analyse some specific solutions in different models. High curvature regime is considered in Section 4. and we make conclusions in Section 5.  

\section{Wormholes in $f(R)$ gravity}

The geometry of a static, spherical symmetric wormhole is given by
\begin{equation}
 ds^{2}=-e^{2\varphi(r)} dt^{2}+\frac{1}{1-\frac{b(r)}{r}}dr^{2}+r^{2}(d\theta^{2}+\sin^{2}\theta d\phi^{2}),
 \label{metrika}
\end{equation}
where $\varphi(r)$ is the redshift function and $b(r)$ is a shape function \cite{Thorne}. 
Functions $\varphi(r)$ and $b(r)$ are arbitrary functions of the radial coordinate $r$, 
which nonmonotonically decreases from infinity to a minimal value $r_{0}$ in the throat 
and increases to infinity. For the throat position $r=r_{0} \Rightarrow b(r_{0}) = 
r_{0}$ the metric tensor component is singular. Nevertheless, the proper distance must be well 
behaved, from which the following integral must be real and 
regular outside the throat \cite{Thorne}:
\begin{equation}
 l(r)=\pm\int^{r}_{r_{0}}\frac{dr}{\sqrt{1-b(r)/r}},
\end{equation}
from which follows the condition:
\begin{equation}
 1-b(r)/r\geq0 .
\end{equation}
So, far from the throat in both radial directions space must be asymptotically flat 
which implies the condition $ b(r)/r\rightarrow0 $ as $ l\rightarrow\pm\infty $ 
\cite{Thorne}.
One of the fundamental wormhole properties is that by definition $b(r)$ must fulfil the 
flaring-out condition at or near the throat: $(b(r)-b(r)'r)/b^{2}>0$ \cite{Thorne}, where 
$b'(r)=db/dr$, (in further text prime denotes a derivative with respect to the argument 
of a function). The second condition which we impose is practical:
we demand that a wormhole must be traversable which means the absence of horizons. So $\varphi(r)$ must be finite everywhere.
Using standard mathematical procedure from (\ref{metrika}) we obtain the Ricci curvature scalar: 
\begin{dmath}
   R=-\frac{2}{r^{2}}[(\varphi(r)''r^{2}+2{\varphi(r)'}^{2}r^{2})(1-\frac{b(r)}{r})   - 
   \frac{\varphi(r)'}{2}(b(r)'r-b(r))  -{\varphi(r)'}^{2}r^{2}(1-\frac{b(r)}{r})     +2\varphi(r)'r(1-\frac{b(r)}{r})-
   r(\frac{b(r)'}{r}-\frac{b(r)}{r^{2}})-\frac{b(r)}{r}].
   \label{ricci}
\end{dmath}  
While studying wormholes in $f(R)$ modified theories of gravity, in order to simplify 
equations, it is common to place an additional condition on red-shift function 
$\varphi(r)$ to be constant \cite{lobo, azizi, Myrzakulov}. This condition on $\varphi(r)$, 
which is assumed without any physical reason, is not justified because the fundamental 
parameters of a wormhole should not be restricted in such an artificial way. Moreover, 
wormhole solutions of modified Einstein's equations and the WEC violation will in some 
cases critically depend on $\varphi(r)$.
Matter is described by the stress-energy tensor of the anisotropic perfect fluid: 
\begin{equation}
T_{\mu \nu}=(\rho + p_{t})U_{\mu}U_{\nu} + p_{t}g_{\mu \nu} + (p_{r} - p_{t})\chi_{\mu}\chi_{\nu}, 
\end{equation}
where $U$ is a four-velocity, $\rho$ the energy density,  $p_{t}$ and $p_{r}$ are transversal and radial pressure respectively, and $\chi^{\mu}= \sqrt{1-b(r)/r}\delta^{\mu}_{r}$.
In (\ref{akcija}) we select $k=1$ for simplicity, and from (\ref{einstein tensor}) we obtain modified Einstein's equations for the wormhole geometry
\begin{dmath}
\frac{b'(r)f_{R}}{r^{2}}= - f_{RR}(1-\frac{b(r)}{r}) 
R'(r)\varphi'(r)+\frac{1}{6}(R(r)f_{R} + f) \\ 
+ \frac{1}{3}(2\rho + 2p_{t} + p_{r}),
\label{prva}
\end{dmath}
\begin{dmath}
\frac{-b(r)+ 2r^{2}\varphi'(r)(1-b(r)/r)}{r^{3}}f_{R} = 
f_{RR} [R''(r)(1-\frac{b(r)}{r}) + \frac{R'(r)}{2}(\frac{b(r)}{r^{2}} - \frac{b'(r)}{r})] 
+f_{RRR}R(r)^{'2}(1-\frac{b(r)}{r}) - \frac{1}{6}(R(r)f_{R}+f) \\+ \frac{1}{3}(\rho + 2p_{r} - 2p_{t}), 
\end{dmath}
\begin{dmath}
(1-\frac{b(r)}{r})(\varphi''(r) - \frac{b'(r)r - b}{2r(r-b)}\varphi'(r) + \varphi'(r)^{2} 
+ \frac{\varphi'(r)}{r} -\frac{b'(r)r - b}{2r^{2}(r-b)})f_{R} = \frac{f_{RR}}{r}(1-\frac{b(r)}{r})R'(r) 
- \frac{1}{6}(R(r)f_{R}+f)+ \frac{1}{3}(\rho - p_{r} + p_{t}).
\label{treca}
\end{dmath}
Note that field equations (\ref{prva} - \ref{treca}) are forth order nonlinear 
differential equations in $\varphi(r)$ and $b(r)$. However, equations (\ref{prva} - 
\ref{treca}) at the same time represent the system of algebraic equations for the stress-energy tensor 
components that, despite it's complexity, have analytic solutions. In our work in 
specific models of modified  $f(R)$ gravity
we consider solutions for the components of the stress-energy tensor by exploring 
different 
redshift, $\varphi(r)$, and shape functions, $b(r)$.
From field equations (\ref{prva} - \ref{treca}) we can derive a specific form
of the WEC for wormhole solutions in $f(R)$ gravity
\begin{dmath}
 \rho=\frac{1}{2 r^{2}} \left[ 2f_{RRR}{R'}^{2}(r)r(b(r)-r) +f_{RR}(2R''(r)r((b(r)-r) \\
 +R'(r)(3b(r)+b'(r)r-4r)) + f_{R}(4r(r-b(r))({\varphi}''(r) \\  
 +{\varphi'}^{2}(r)) - 2b'(r)(1+\varphi'(r)) + 2\varphi'(r) (4r-3b(r)) \\
 +r^{2}R(r))+r^{2}f(R)\right] \geq0,
 \label{wec1}
\end{dmath}
\begin{dmath}
 \rho+p_{r}=\frac{1}{ r^{3}} \left[ f_{RRR}{R'}^{2}(r)r^{2}(b(r)-r) + f_{RR}(r^{2}R''(r)(b(r)-r) \\
 -R'(r) r^{2}\varphi'(r)(b(r)-r) + \frac{R'(r)}{2}r(b'(r)r-b(r))  \\
 +f_{R}((b'(r)r-b(r))  -2\varphi'(r) r(b(r)-r)) \right] \geq0.
 \label{wec2}
\end{dmath}
\begin{dmath}
 \rho+p_{t}=\frac{1}{ r^{2}} \left[ [ \frac{b'(r)r+b(r)}{2r} - (b(r)-r)(r\varphi''(r) \\
 + r {\varphi'}^{2}(r)+\varphi'(r)) -\frac{b'(r)r-b(r)}{2}\varphi'(r)]f_{R} \\
   + f_{RR} R'(r) (b(r)-r) (1-r\varphi'(r))   \right] \geq0.
 \label{wec3}
\end{dmath}
We require that the matter threading the wormhole satisfies WEC, so we demand that 
inequalities (\ref{wec1}),(\ref{wec2}) and (\ref{wec3}) are fulfilled. Since in Einstein's general
relativity, which corresponds to $f(R)=R$, this is not possible, higher 
curvature terms in the action support wormhole geometries. We can see that 
explicit analysis of the equations (\ref{wec1}),(\ref{wec2}) and (\ref{wec3})  is extremely 
difficult, and that for a specific wormhole geometry WEC violation can critically 
depend on the redshift functions $\varphi(r)$ and its derivatives. As an important 
and interesting case we can consider the equations (\ref{wec2}) and (\ref{wec3}) near the throat.
This approach simplifies the problem considerably. Near the throat
 $b(r)\simeq r_{0}$ and the 
equations (\ref{wec2}) and (\ref{wec3}) become
\begin{equation}
\rho + p_{r}=\frac{b'(r)r-b(r)}{ 2r^{3}}[f_{RR}R'(r)r+2f_{R}]\geq0.
\label{pavsosspr}
\end{equation}
\begin{equation}
\rho + p_{t}=\frac{f_{R}}{r^{2}} \left[ \frac{b'(r)r+b(r)}{ 2r}- \frac{b'(r)r-b(r)}{ 2}\varphi'(r) \right] \geq0.
\label{pavsosspt}
\end{equation}
From the flaring-out condition we must have $b'(r)r-b(r)<0$,
so for equation (\ref{pavsosspr}) we simply obtain 
\begin{equation}
f_{RR}R'(r)r+2f_{R}\leq0 ~~~\textrm{near the throat}.
\label{pavlsoss}
\end{equation}
This condition is, due to its simplicity, particularly suitable for analysing
the influence of modifying the theory of gravity on the question of the WEC violation.
It is obvious that this condition cannot be fulfilled for every choice of $f(R)$.
For instance, if we take $f(R)=R$ this condition is not satisfied and this 
corresponds to the need for exotic matter in Einstein's relativity.
As we will show in the later part of our work, for analysed models we typically have $f_{R}<0$ near the throat, and from (\ref{pavsosspt}) we obtain the condition
\begin{equation}
 b'(r) \leq \frac{b(r)(1+r \varphi'(r))}{ r^{2}\varphi'(r) - r}.
\end{equation}
If $\varphi(r)$ is taken to be differentiable and continuous function on the interval $[r_{0}, \infty\rangle$ and $\frac{(1+r \varphi'(r))}{ r^{2}\varphi'(r) - r}$ 
continuous on the same interval, then according to Gronwall-Bellman inequality we have
\begin{equation}
 b(r) \leq b(r_{0})\exp \left( \int_{r_{0}}^{r} \frac{1+z\varphi'(z)}{z(z\varphi'(r)-1)} dz \right).
 \label{integral}
\end{equation}
From this expression we can see the importance of $\varphi(r)$:  for a given red-shift function we can solve the integral (\ref{integral}) and determine the 
function that bounds $b(r)$ near the throat. Taking this bounding function as a critical case we can determine Ricci scalar and check whether
(\ref{pavsosspr}) is satisfied in the concrete $f(R)$ model. 
Let us consider a specific example $\varphi(r)=constant$ for simplicity. Bounding shape function is then
\begin{equation}
 b(r)_{critical}= b(r_{0})\frac{r_{0}}{r}.
\end{equation}
Then, calculating the Ricci scalar and its derivative, from (\ref{pavsosspr}) we obtain the condition: 
\begin{equation}
\frac{f_{R}}{f_{RR}} \leq - \frac{4 b(r_{0})r_{0}}{r^{4}},
\end{equation}
which depends only on a specific $f(R)$ model when the constants $b(r_{0})$ and $r_{0}$ are given.
\\
 
\begin{figure}[h]
  \centering
  \includegraphics[scale=0.25]{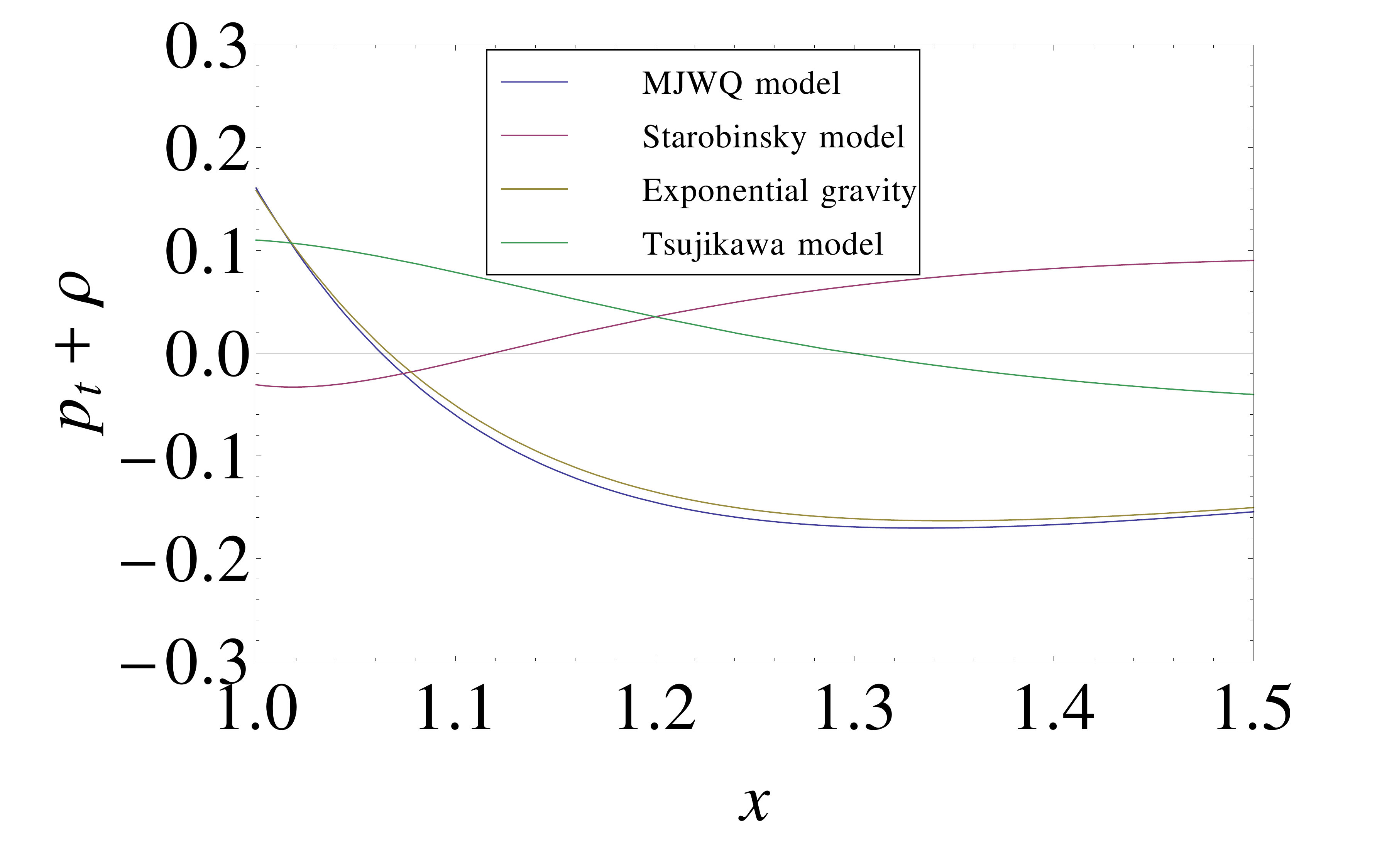}
  \caption{$\rho+p_{t}$ for all models respectively, MJWQ model: $b(r)=r_{0}\sqrt{r_{0}/r}$, $\varphi(r)=\ln(r_{0}/r+1)$, and parameters $\beta=2$, $R_{*}=2$,
  The Starobinsky model: $b(r)=r_{0}\ln{r}/r_{0}+r_{0}$, $\varphi(r)=\sqrt{r_{0}/r}$ and parameters $q=2$, $\lambda=1$,
  The Exponential gravity model: $b(r)=r_{0}\sqrt{r_{0}/r}$, $\varphi(r)=\ln(r_{0}/r+1)$, and parameters $\lambda=2$, $R_{*}=2$, 
  The Tsujikawa model: $b(r)=r_{0}\sqrt{r_{0}/r}$, $\varphi(r)=\ln(r_{0}/r+1)$, and parameters $\mu=2$ $R_{*}=1$, where $x=r/r_{0}$.}
  \label{wec3zasve}
\end{figure}

 In the next section our approach will be to find specific solutions of equations 
(\ref{prva} - \ref{treca}) for a given wormhole geometry and then check whether the WEC 
is satisfied or not, rather than analyse equations (\ref{wec1}), (\ref{wec2}) and (\ref{wec3}). We will focus on the question of 
condition $(\ref{wec2})$ violation for the specific $f(R)$ models, since it is this part of the WEC that is necessary violated near the throat in the 
Einstein's general theory of relativity, and was used as a definition for exotic matter in \cite{Thorne}. For the simple choices of $b(r)$ and $\varphi(r)$ in 
the considered models, $\rho + p_{t}$ is typically violated somewhere away from the throat, as shown in the Fig. \ref{wec3zasve} for all models. 
This should not be of primary concern, since it is the throat connecting two asymptotically flat regions that is of the main interest, and one can always cut-off the solution 
at some $r_{c}$ away from the throat and connect it with other asymptotically flat solution of modified Einstein's equations in that region. This would physically correspond to 
a wormhole in a spacetime in which another energy-momentum distribution starts to dominate for $r\geq r_{c}$.

\section{Specific models and solutions}

\begin{figure}[h]
  \centering
  \includegraphics[scale=0.25]{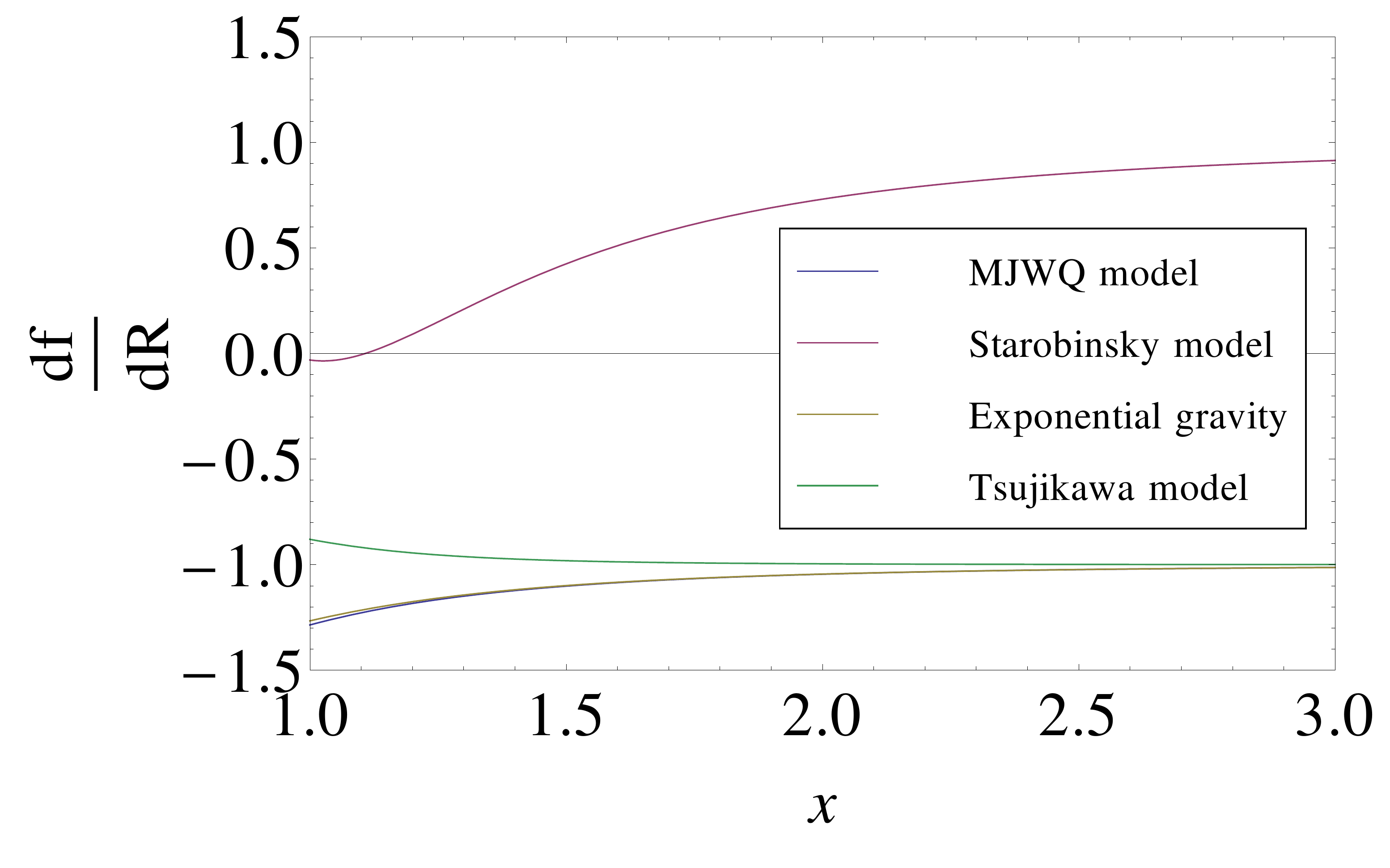}
  \caption{The MJWQ model: $b(r)=r_{0}\sqrt{r_{0}/r}$, $\varphi(r)=\ln(r_{0}/r+1)$, and parameters $\beta=2$, $R_{*}=2$,
  The Starobinsky model: $b(r)=r_{0}\ln{r}/r_{0}+r_{0}$, $\varphi(r)=\sqrt{r_{0}/r}$ and parameters $q=2$, $\lambda=1$,
  The Exponential gravity model: $b(r)=r_{0}\sqrt{r_{0}/r}$, $\varphi(r)=\ln(r_{0}/r+1)$, and parameters $\lambda=2$, $R_{*}=2$, 
  The Tsujikawa model: $b(r)=r_{0}\sqrt{r_{0}/r}$, $\varphi(r)=\ln(r_{0}/r+1)$, and parameters $\mu=2$ $R_{*}=1$, where $x=r/r_{0}$.}
  \label{fRzasvemodele}
\end{figure}

In some works which analyse wormholes in the context of $f(R)$ gravity \cite{lobo, 
azizi} $f(R)$ is usually treated as an unknown function, which can be derived from  
modified field equations, or it is considered to have some simple convenient shape. We prefer the approach in which $f(R)$ functions are taken as predetermined 
characteristic of the theory. In fact, due to the highly hypothetical nature of a 
wormhole, which is at the moment far away from any empirical observation, we cannot 
impose conditions on $f(R)$ in a manner stated above. Therefore, the form of $f(R)$ should be 
consistent with observations for the Solar System and cosmological evolution, so 
we analyse wormhole solutions in several viable models of $f(R)$ gravity
 \cite{Jaime, salgado, 
linder, motohashi, odintsov}:
 \begin{itemize}
 \item MJWQ model \cite{miranda}
 \begin{equation}
 f(R)=R- \beta R_{*}\ln(1+\frac{R}{R_{*}}),
 \label{mjw}
 \end{equation}
where  $\beta$ and $R_{*}$ are free positive parameters of the model.
 \item Starobinsky model \cite{starobinsky, amendola, tsu, polarski, gan, apt}
 \begin{equation}
 f(R)=R + \lambda R_{*}[(1+ ( \frac{R^{2}}{R_{*}^{2}}))^{- q}-1],
 \end{equation}
 with three free positive parameters $\lambda$, $R_{*}$ and $q$.
\item Exponential gravity model \cite{cognola, elizalde}
\begin{equation}
f(R)= R - R_{*}\lambda(1 - e^{-\tilde{R}}),
 \end{equation}
 where $\tilde{R}=R/R_{*}$, with $\lambda$ and $R_{*}$ as free positive parameters of the model. 
 \item Tsujikawa model \cite{tsu, felice}
 \begin{equation}
 f(R)=R-\mu R_{*} \tanh (\frac{R}{R_{*}})
 \end{equation}
 where $\mu$ and $R_{*}$ are free positive parameters of the model.
\end{itemize} 
In all models $R_{*}=\sigma H_{0}^{2}$, where $\sigma$ is some dimensionless parameter and $H_{0}$ is the current value of the Hubble parameter, which is taken to be $H_{0}=1$.
\\ \\
It was shown in \cite{starobinskiteorem1}, \cite{starobinskiteorem2} that for a scalar-tensor theory of gravity, formulated in the Jordan frame with the field Lagrangian
\begin{equation}
L=\frac{1}{2}[f(\Phi)R + h(\Phi)g^{\mu \nu}\Phi_{, \mu} \Phi_{, \nu} - 2U(\Phi)],
\end{equation}
no static wormholes that satisfy the Null Energy Condition can be formed, as long as $f(\Phi)$ is everywhere positive and also
\begin{equation}
f(\Phi)h(\Phi)+\frac{3}{2}(\frac{d f}{d \Phi})^{2} >0, 
\end{equation}
where $f$, $h$, and $U$ are arbitrary functions. For $f(R)$ gravity we have $f(\Phi)=f_{R}$, $h=0$ and $2U(\Phi)=Rf_{R} - f(R)$. Therefore, two conditions 
for wormholes non-existence in $f(R)$ gravity read
$f_{R} >0$ and $f_{RR}>0$. As can be seen from the Fig. \ref{fRzasvemodele} the considered models have regions with $f_{R}<0$ and therefore violate the non-existence 
theorem conditions. However, this opens the question of the stability of solutions under non-static perturbations which should be considered in the further work.

\subsection{MJWQ model}

\begin{figure}[h]
  \centering
  \includegraphics[scale=0.25]{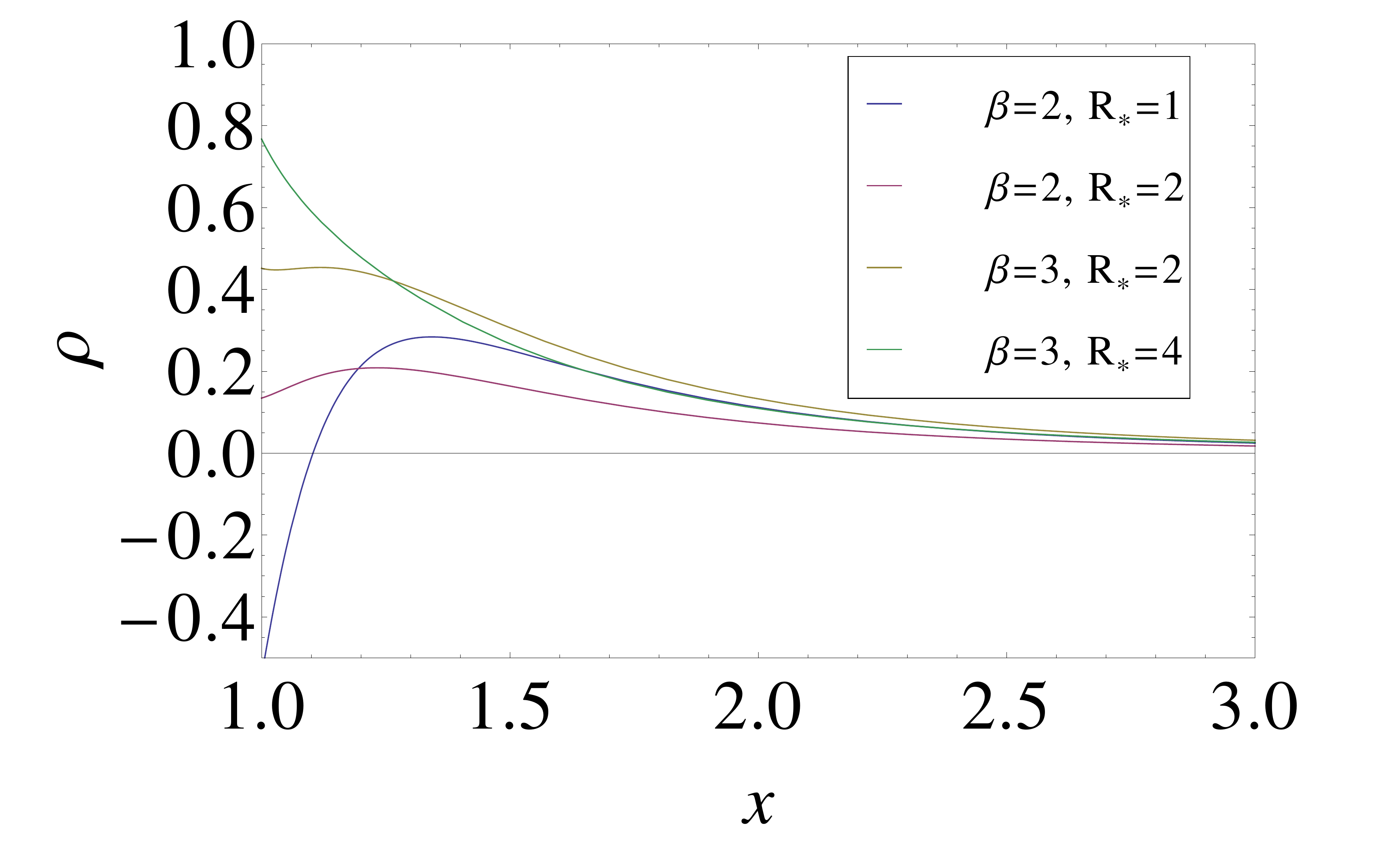}
  \caption{Energy density, $\rho$, in the MJWQ model for the specific choice 
  of $b(r)=r_{0}\sqrt{r_{0}/r}$, $\varphi(r)=\ln(r_{0}/r+1)$, where $x=r/r_{0}$.}
  \label{mjw1}
  \end{figure}

\begin{figure}[h]
  \centering
  \includegraphics[scale=0.25]{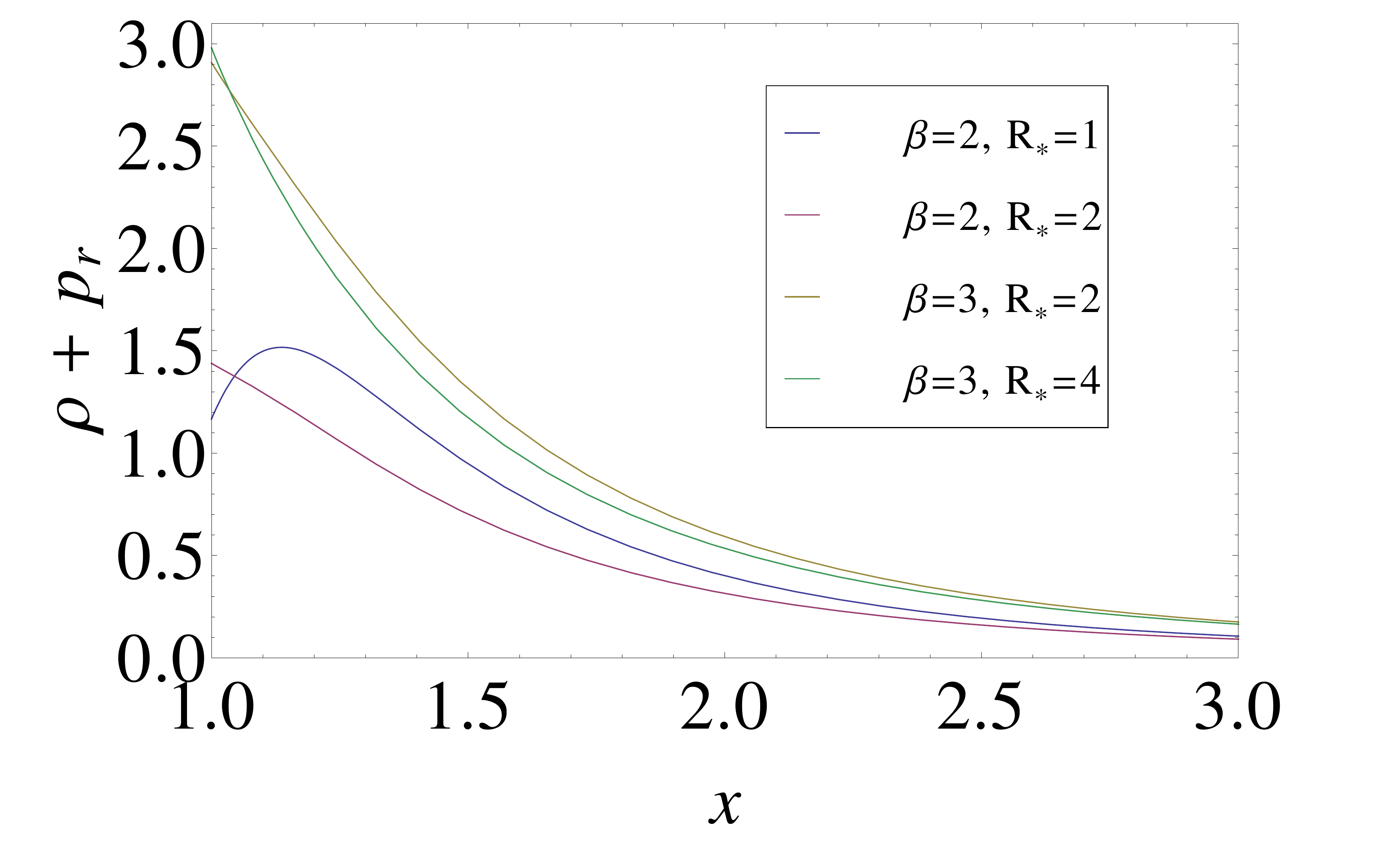}
  \caption{The non-exotic material condition, $\rho+p_{r}$, in the MJWQ model for the specific choice of $b(r)=r_{0}\sqrt{r_{0}/r}$, $\varphi(r)=\ln(r_{0}/r+1)$, where $x=r/r_{0}$.}
  \label{mjw2}
  \end{figure}

In MJWQ model (\ref{mjw}) we solve field equations (\ref{prva} - \ref{treca}) to obtain components of the stress-energy tensor and check 
if the conditions $\rho\geq0$ and $\rho+p_{r}\geq0$ are satisfied.
We consider specific red-shift and shape functions given by  $\varphi(r)=\ln(r_{0}/r+1)$ and $b(r)=r_{0}\sqrt{r_{0}/r}$.
Parameters of the model, $\beta$ and  $R_{*}$,  are taken to be close to the values proposed in \cite{Jaime}. Above mentioned solutions are depicted in Fig. \ref{mjw1} and Fig. \ref{mjw2}. A choice of the 
free parameters in the $f(R)$ model plays a significant role in the question of the WEC violation.
For all shown combinations of parameters both conditions are satisfied, except the case $\beta=2$ and $R_{*}=1$, which was proposed in \cite{Jaime}.

\subsection{Starobinsky model}

\begin{figure}[h]
  \centering
  \includegraphics[scale=0.25]{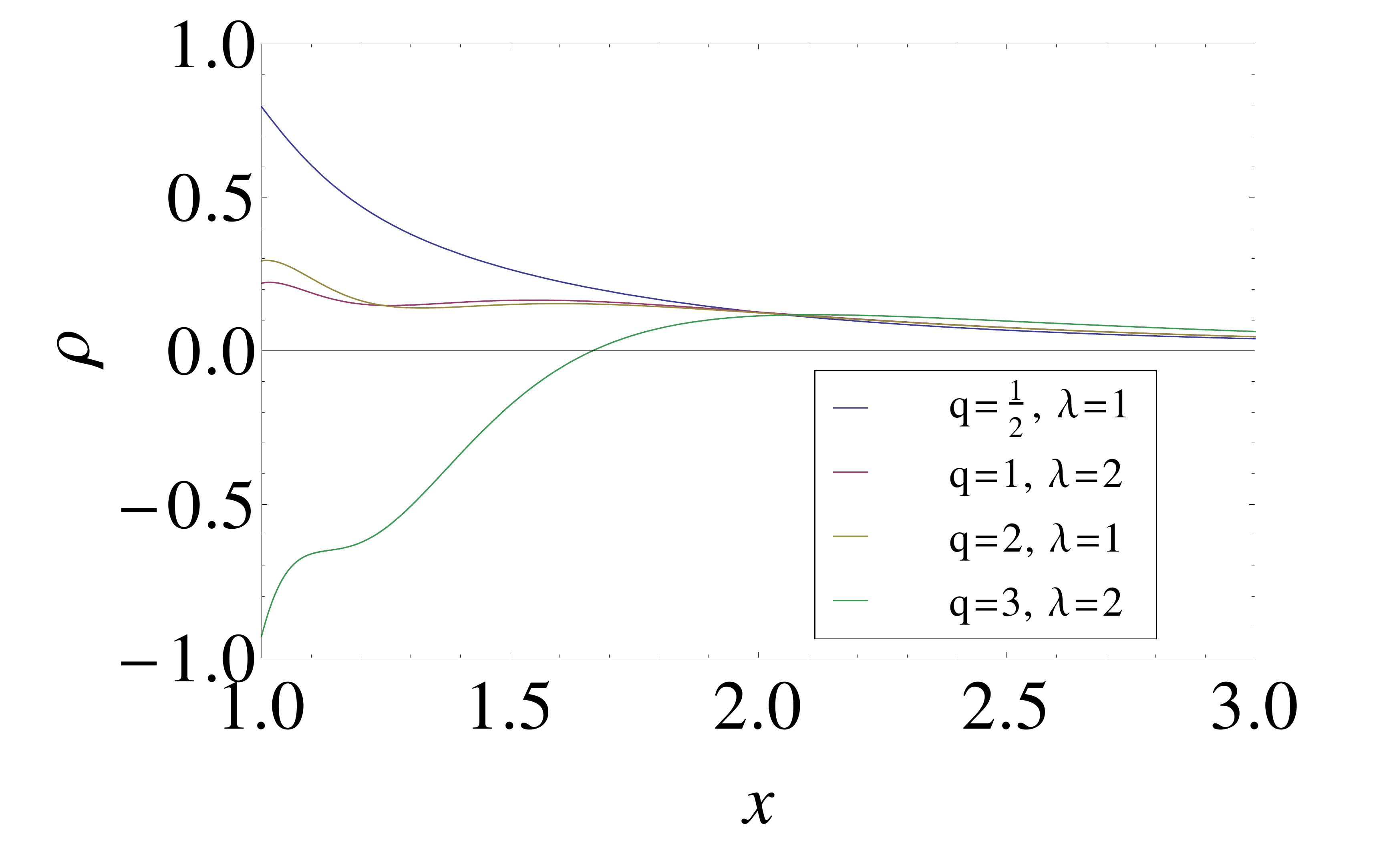}
  \caption{Energy density, $\rho$, in the Starobinsky model for the specific choice 
  of $b(r)=r_{0}\ln{r}/r_{0}+r_{0}$, $\varphi(r)=\sqrt{r_{0}/r}$, where $x=r/r_{0}$.}
  \label{str1}
  \end{figure}

\begin{figure}[h]
  \centering
  \includegraphics[scale=0.25]{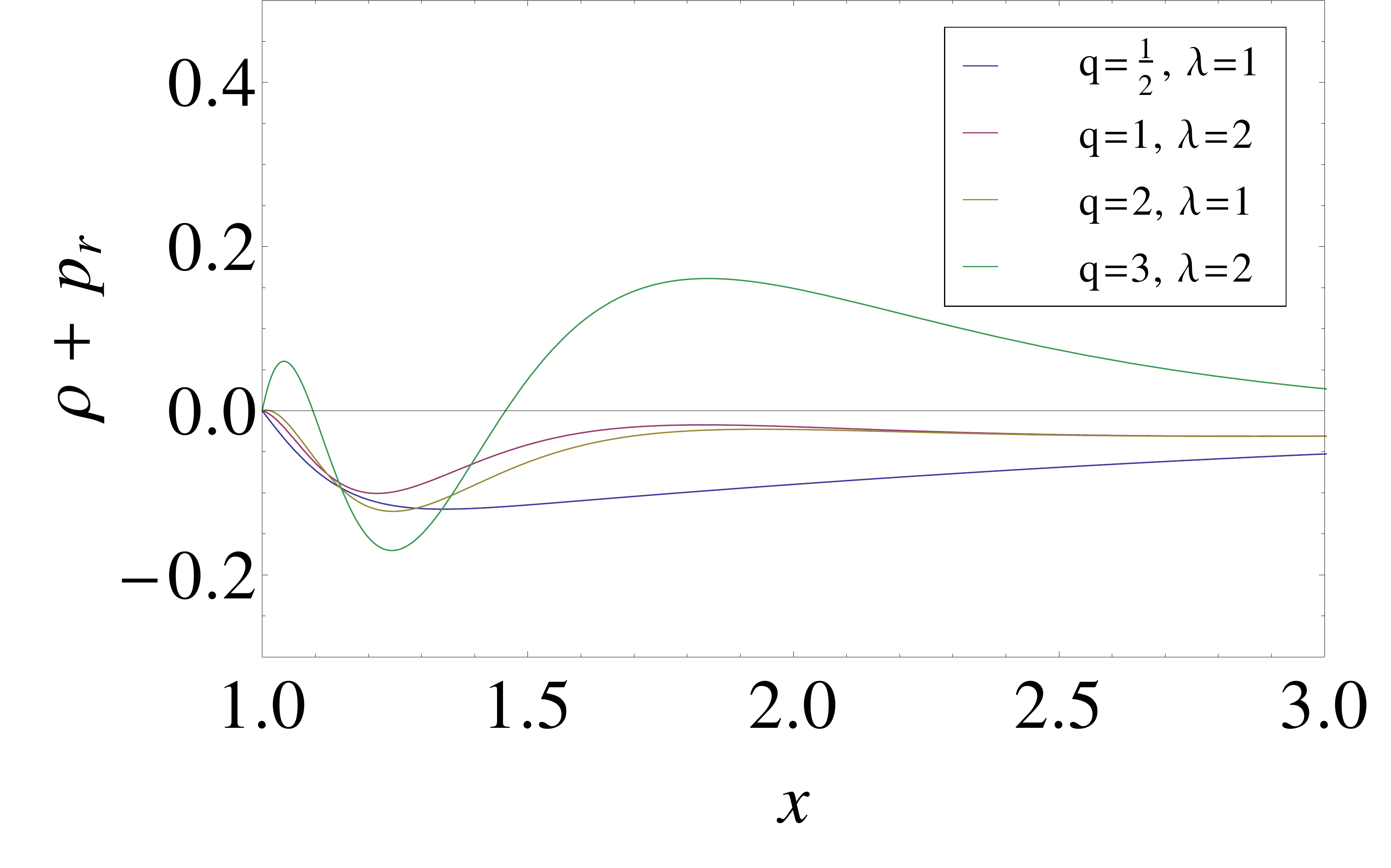}
  \caption{The non-exotic material condition, $\rho+p_{r}$, in the Starobinsky model for the specific choice of $b(r)=r_{0}\ln{r}/r_{0}+r_{0}$, $\varphi(r)=\sqrt{r_{0}/r}$, where $x=r/r_{0}$.}
  \label{str2}
  \end{figure}

Let us consider specific functions $b(r)=r_{0}\ln{r}/r_{0}+r_{0}$, $\varphi(r)=\sqrt{r_{0}/r}$.
As in \cite{Jaime} we choose $R_{*}=4.17$ with $\lambda$ and $q$ close to the values $\lambda=1$, $q=2$. The solutions are depicted in Fig. \ref{str1} and Fig. \ref{str2}. We see that every combination of parameters implies the need for exotic matter. Moreover, 
for every considered combination of simple shape and red-shift functions we did not find non-exotic matter solutions in the Starobinsky model.

  \subsection{Exponential gravity model}
  
\begin{figure}[h]
  \centering
  \includegraphics[scale=0.25]{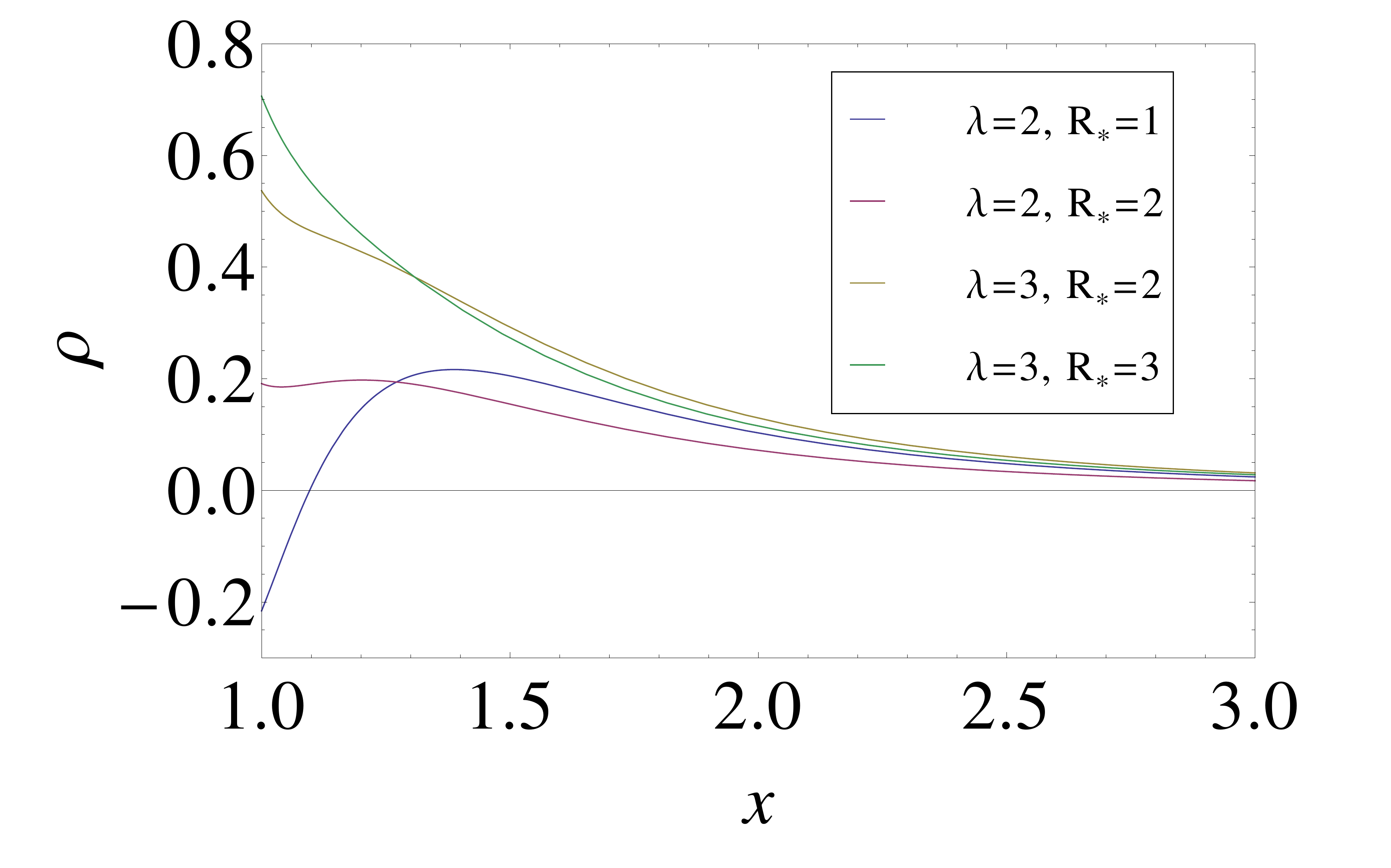}
  \caption{Energy density, $\rho$, in the Exponential gravity model for the specific choice 
  of $b(r)=r_{0}\sqrt{r_{0}/r}$, $\varphi(r)=\ln(r_{0}/r+1)$, where $x=r/r_{0}$.}
  \label{eks1}
  \end{figure}

\begin{figure}[h]
  \centering
  \includegraphics[scale=0.25]{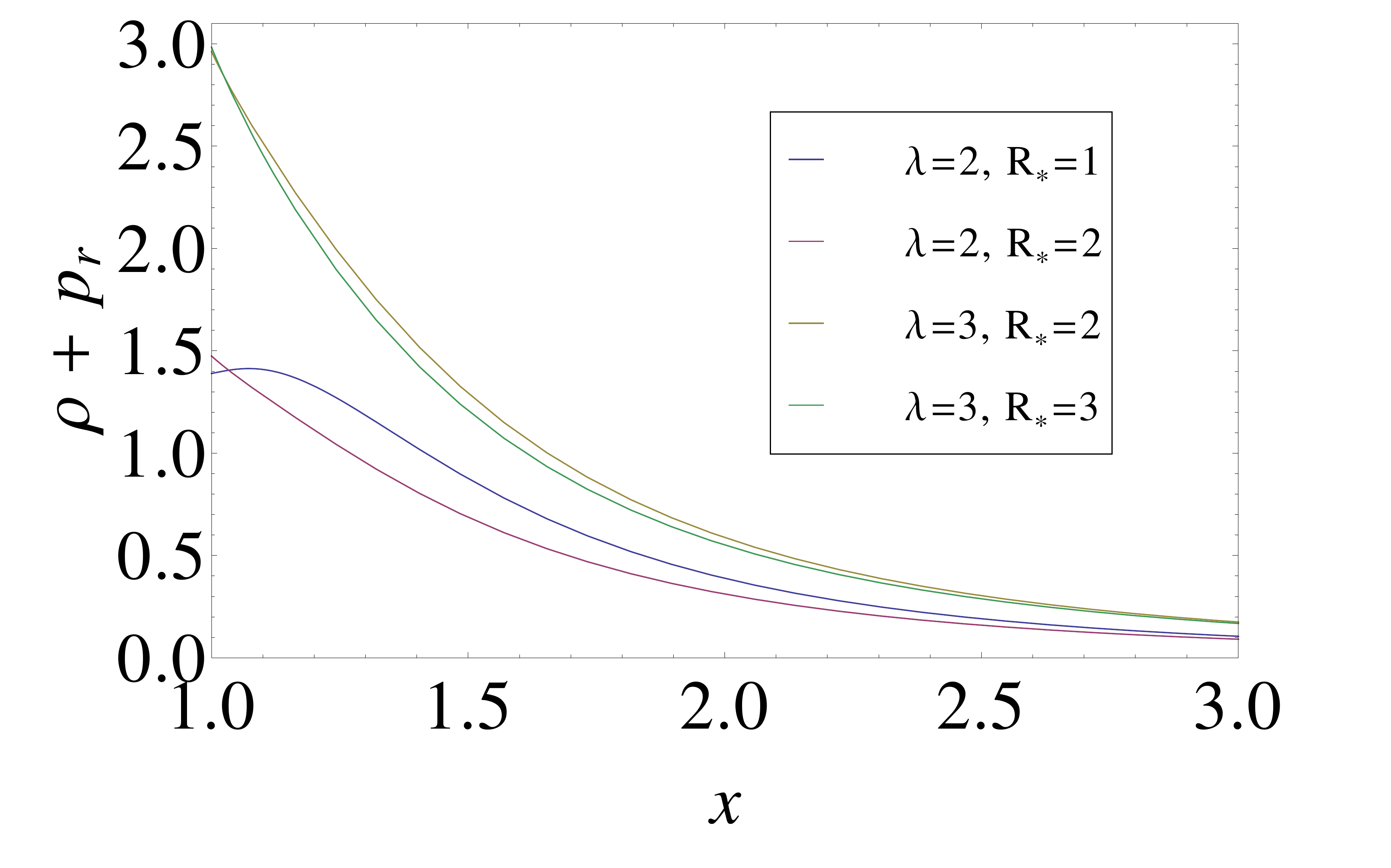}
  \caption{The non-exotic material condition, $\rho+p_{r}$, in the Exponential gravity model for the specific choice of $b(r)=r_{0}\sqrt{r_{0}/r}$, $\varphi(r)=\ln(r_{0}/r+1)$, where $x=r/r_{0}$.}
  \label{eks2}
  \end{figure}

 We take the shape and the redshift functions previously considered in the MJWQ model with $\lambda$ and $R_{*}$ close to the values in \cite{Jaime}. 
Solutions are presented in Fig. \ref{eks1} and Fig. \ref{eks2}.
For all combinations the conditions are satisfied except for the choice $\lambda=2$ and $R_{*}=1$.
  
  \subsection{Tsujikawa model}
  
\begin{figure}[h]
  \centering
  \includegraphics[scale=0.25]{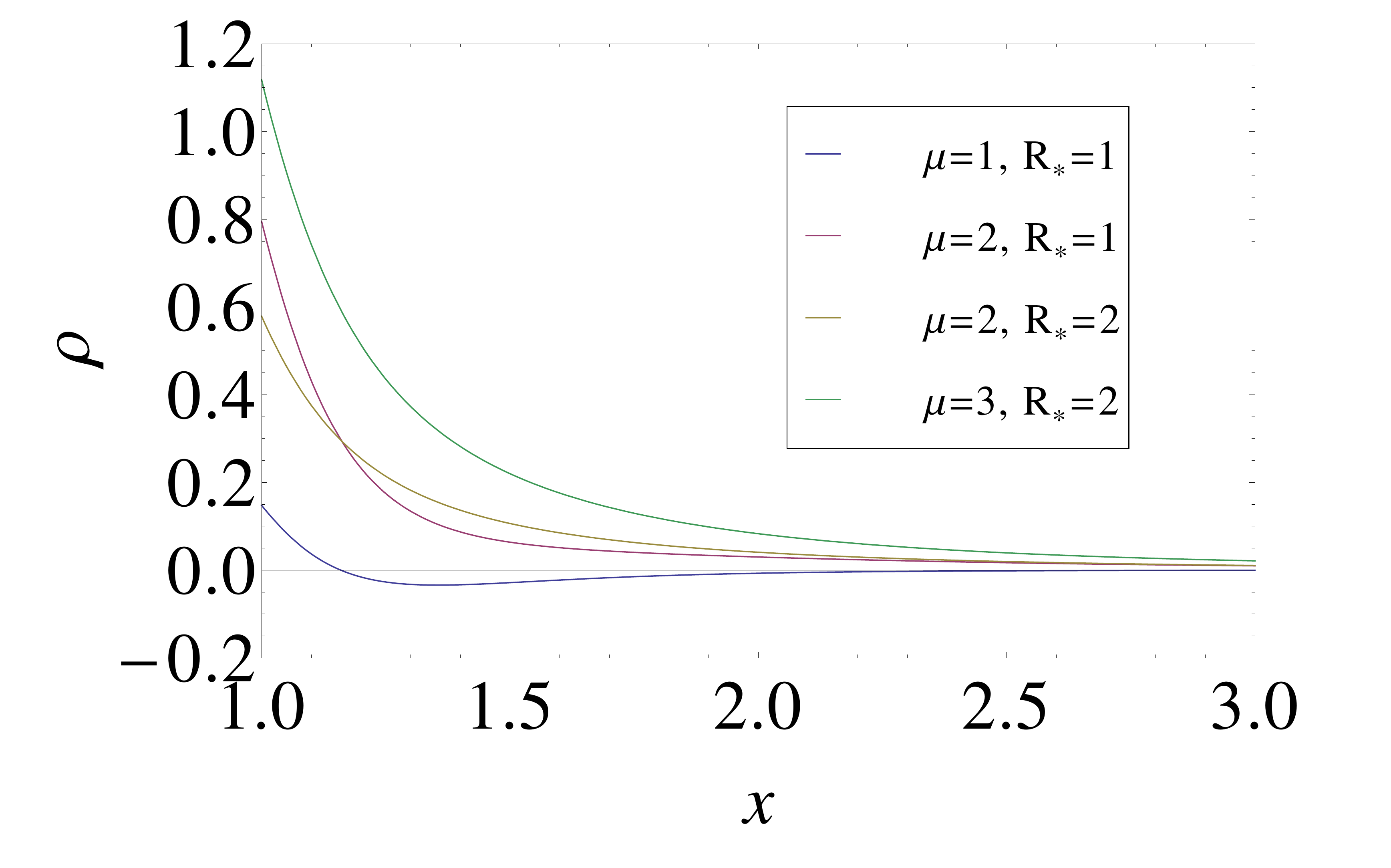}
  \caption{Energy density, $\rho$, in the Tsujikawa
 model for the specific choice 
  of $b(r)=r_{0}\sqrt{r_{0}/r}$, $\varphi(r)=\ln(r_{0}/r+1)$, where $x=r/r_{0}$.}
  \label{tsu1}
\end{figure}

\begin{figure}[h]
  \centering
  \includegraphics[scale=0.25]{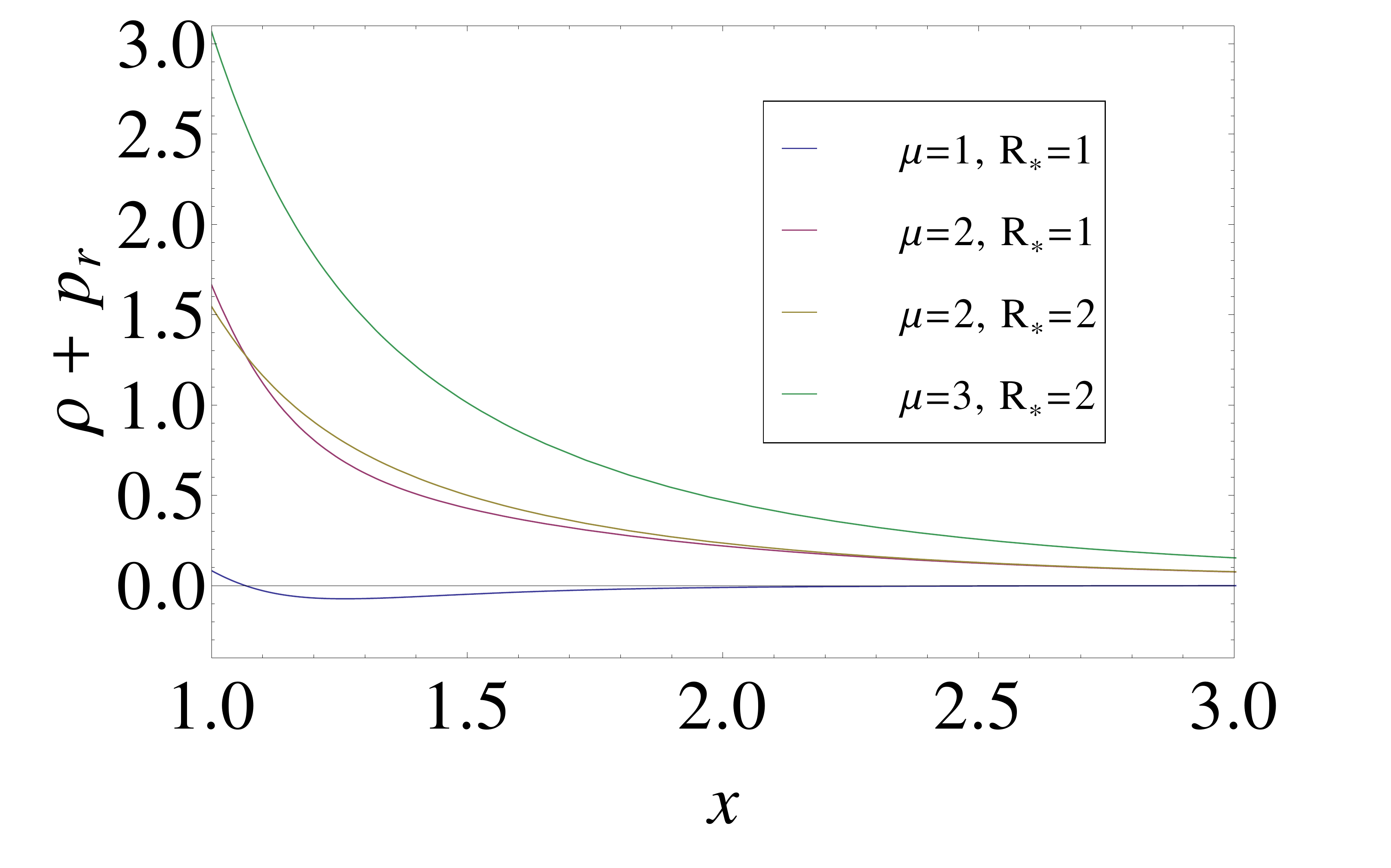}
  \caption{The non-exotic material condition, $\rho+p_{r}$, in the Tsujikawa
 model for the specific choice of $b(r)=r_{0}\sqrt{r_{0}/r}$, $\varphi(r)=\ln(r_{0}/r+1)$, where $x=r/r_{0}$.}
  \label{tsu2}
\end{figure}

Finally, the results for the Tsujikawa model are plotted in Fig. \ref{tsu1} and Fig. \ref{tsu2}. For comparison we choose the same shape and red-shift 
functions as in the MJWQ and the Exponential gravity model. For values smaller than $\mu=2$ and $R_{*}=1$ the model demands the need for exotic matter. 
  
\section{High curvature regime}

It is interesting to consider the high curvature limit in the problem
 of wormholes in viable $f(R)$ modified theories of gravity. Specific $f(R)$
 models which are considered in this work should reproduce realistic scenarios 
 of cosmological evolution based on the accelerated expansion of the universe.
  Cosmological models based on viable $f(R)$ theories asymptotically approach
  the de-Sitter solution which in Einstein's general relativity corresponds to an
  empty space filled with the positive cosmological constant \cite{amendola2}.
  In accordance with this demand all considered models share the same
  mathematical property that in the high curvature limit they have the 
  following form
  \begin{equation}
  f(R)\approx R-\lambda R_{*},
  \label{curvature}
  \end{equation}
  so the models lead to the effective cosmological constant  $\lambda R_{*}/2$,
  as can be seen by applying (\ref{curvature}) to (\ref{modificirana 
  einsteinova}) and using the standard definition of the cosmological constant in 
   Einstein's field equations. Therefore the high curvature regime is interesting 
  because we have an interplay of cosmological features of the $f(R)$ models and
   wormhole solutions. We expect to have the high curvature limit in the vicinity 
   of
  the throat for a suitable choice of $b(r)$ and $\varphi(r)$ which will lead to 
  $R(r)\gg R_{*}$. By rewriting Einstein's field equations (\ref{prva}-\ref{treca}), with 
  (\ref{ricci}) and (\ref{curvature}), we can easily obtain the solutions for   
  the stress-energy tensor
  components in the high curvature limit
  \begin{equation}
   \rho=-\frac{-2b'(r)+\lambda R_{*}r^{2}}{2r^{2}},
  \end{equation}
  \begin{equation}
   p_{r}=-\frac{2 b(r) + 4 b(r) \varphi'(r) r - 4 \varphi'(r) r^2 -\lambda R_{*} r^3}{2r^{3}},
  \end{equation}
\begin{dmath}
   p_{t}=\varphi''(r) + \varphi(r)^2 + \frac{\lambda R_{*}}{2} + \frac{b(r)}{2 r^3} - \frac{b'(r)}{2 r^{2}} \\
   -    \frac{b(r)\varphi'(r)}{ 2 r^2} - \frac{b(r)\varphi''(r)}{r} + \frac{\varphi'(r)}{r} \\ 
   - \frac{b'(r)\varphi'(r)}{ 2 r} - \frac{b(r)\varphi'(r)^{2}}{r}.
\end{dmath}
It can be seen that these equations are equal to the ones presented in \cite{Thorne} with
  addition of an effective cosmological constant, as should be expected. 
  By inspecting (\ref{pavlsoss}) it is apparent that the WEC is violated near the throat
  in the high curvature regime. Since all viable $f(R)$ models share the same 
  asymptotic behaviour described by (\ref{curvature}) one can question  
  the critical $R(r)$ value for every point near the throat, in a specific $f(R)$
  model. By critical value we mean the highest possible $R(r)$ value for which WEC is
  satisfied at a specific point in space. Let us consider solutions, $R_{critical}(r)$, of the following
  equation that can be obtained from (\ref{pavlsoss})
  \begin{equation}
   f_{RR}R_{critical}'(r)r+2f_{R}=0 ~~~\textrm{near the throat},
   \label{pavlsoss2}
  \end{equation}
  as well as solutions, $\bar{R}$, of the WEC violation inequality
  \begin{equation}
   f_{RR}\bar{R}'(r)r+2f_{R}>0.
   \label{usporedba1}
  \end{equation}
  From the theory of differential inequalities follows
  \begin{equation}
   \bar{R}(r)>R_{critical}(r),
   \label{usporedba2}
  \end{equation}
  in the interval near the throat $r_{0}<r<r_{1}$, where
  $\bar{R}(r_{0})=R_{critical}(r_{0})$. Therefore, values of $R_{critical}$ 
  correspond to the critical values of the Ricci scalar in the above mentioned 
  sense.
  We can solve the equation (\ref{pavlsoss2}) and obtain $R_{critical}$ in different 
  models of $f(R)$
  gravity. For instance in the MJWQ model we get
  \begin{equation}
    R_{critical}(r)=\frac{(\frac{r^{2}}{r_{0}^{2}})^{2}-\frac{1-R_{0}}{1+R_{0}}}{(\frac{r^{2}}{r_{0}^{2}})^{2}+\frac{1-R_{0}}{1+R_{0}}},
  \end{equation}
  where $R_{*}=1$, $\beta=2$, $R(r_{0})=R_{0}$. 
  In Exponential gravity for fixed parameters $q=2$, $\lambda=2$ we obtain
  \begin{equation}
   R_{critical}(r)=\ln \frac{4r^{2}}{(c+r^{2})^{2}},
  \end{equation}
  with $c=\frac{2r_{0}^{2}}{e^{R_{0}/2}}-r_{0}^{2}$. In this way it is possible
  to  considerably simplify the analysis of the wormhole WEC violation in $f(R)$ theories of gravity. 
  For a given $R(r)$ one can compare its values near the throat with values of $R_{critical}(r)$
  in a concrete $f(R)$ model and using (\ref{usporedba2}) and (\ref{usporedba1}) check whether WEC is violated.
  Of course, non violation of (\ref{pavlsoss}) is necessary, but not sufficient for WEC non violation.  
  
  \section{Conclusions}
  
  We have examined possible wormhole solutions in four viable recently proposed $f(R)$ 
models, namely: the MJWQ model, the Starobinsky model, the Exponential gravity model and 
the Tsujikawa model. In all models apart from the Starobinsky model we have found
 solutions 
that do not require exotic matter. Our solutions do not presume any 
additional assumption on the redshift function, $\varphi(r)$. 
For given functions $\varphi(r)$ and $b(r)$ we can see that for the above cases 
the character of 
the solutions in the MJWQ, Exponential gravity and Tsujikawa model depends more 
strongly on the choice of the free parameters than on the choice of a specific model. 
It is also possible to satisfy non-exotic matter conditions with other simple 
choices of $\varphi(r)$ and $b(r)$ for each model. For instance 
$b(r)=r_{0}e^{(1-r/
r_{0})}$, $b(r)=r_{0}^{2}/r$, $\varphi(r)=1/r$, etc. 

Simple inequality for the WEC violation near the throat is derived, which we demonstrate to be particularly suitable for analysing
the influence of modifying theory of gravity on the question of the WEC violation. We have shown that all viable $f(R)$ models 
must share the same mathematical form in the high curvature regime and that in this limit the WEC is necessarily violated. 
 The question of critical values of the Ricci scalar i.e. 
  the highest possible values of the Ricci scalar for which the WEC is
  not violated near the throat, is considered.  We have  calculated functions of the Ricci scalar that give these 
  critical values for several models. Following this approach, and comparing values of some arbitrary Ricci scalar
  near the throat with critical values, it is straightforward to check whether the necessary condition of the WEC non violation is satisfied.
  While in some previous works WEC violation was analysed in some simple $f(R)$ frameworks we have considered viable and realistic models, and 
  showed that wormholes that do not demand exotic matter can be constructed in them.
In further work it would be 
interesting to analyse stability of the solutions, as well as solutions in non-spherically symmetric wormholes and wormholes 
supported by scalar and gauge fields instead of perfect anisotropic fluid, in $f(R)$ 
theories of gravity.

\begin{acknowledgements}
We would like to thank Professor Dubravko Horvat for fruitful discussions, useful 
comments and constructive suggestions which have improved this work considerably. 
\end{acknowledgements}

\end{document}